\documentclass[usenatbib]{mn2e}		

\usepackage{epsfig}

\begin{document}

\title[Energy-Resolved Photometry of UZ~For]
{High-Speed Energy-Resolved STJ Photometry of the Eclipsing Binary
UZ~For\thanks{Based on observations made with the William Herschel
Telescope operated on the island of La Palma by the Isaac Newton Group
in the Spanish Observatorio del Roque de los Muchachos of the
Instituto de Astrofisica de Canarias}}

\author[M.A.C. Perryman et al.]
{M.A.C.\,Perryman$^{1}$,
M.\,Cropper$^{2}$,
G.\,Ramsay$^{2}$,
F.\,Favata$^{1}$,
A.\,Peacock$^{1}$,
\newauthor
N.\,Rando$^{1}$,
A.\,Reynolds$^{1}$
\\
$^1$Astrophysics Division, Space Science Department of ESA, ESTEC,
  Postbus~299, 2200~AG Noordwijk, The Netherlands\\
$^2$Mullard Space Science Laboratory, University College London,
	Holmbury St~Mary, Dorking, Surrey RH5~6NT, UK
}

\date{Accepted ; Received}
\pubyear{2001}

\maketitle 

\begin{abstract}
We present high time-resolution optical photometry of the eclipsing
binary UZ~For using a superconducting tunnel junction (STJ) device, a
photon-counting array detector with intrinsic energy resolution. Three
eclipses of the $\sim$18~mag 126.5~min orbital binary were observed
using a $6\times6$ array of Tantalum STJs at the 4.2-m William
Herschel Telescope on La~Palma. The detector presently provides
individual photon arrival time accuracy to about 5~$\mu$s, and a
wavelength resolution of about 60~nm at 500~nm, with each array
element capable of counting up to $\sim$5000~photons~s$^{-1}$. The
data allow us to place accurate constraints on the accretion geometry
from our time- and spectrally-resolved monitoring, especially of the
eclipse ingress and egress. We find that there are two small accretion
regions, located close to the poles of the white dwarf. The positions
of these are accurately constrained, and show little movement from
eclipse to eclipse, even over a number of years. The colour of the
emission from the two regions appears similar, although their X-ray
properties are known to be significantly different: we argue that the
usual accretion shock may be absent at the non-X-ray emitting region,
and instead the flow here interacts directly with the white dwarf
surface; alternatively, a special grazing occultation of this region
is required. There is no evidence for any quasi-periodic oscillations
on time-scales of the order of seconds, consistent with relatively
stable cyclotron cooling in each accretion region.
\end{abstract}


\section{Introduction}
\label{sec:intro}

UZ~For is a member of the AM~Herculis type cataclysmic variables
(CVs), in which a strongly magnetic white dwarf, with a polar field
strength of order $5-7\times10^7$G, accretes material from a late-type
companion that fills its Roche lobe (see \citealt{cro90} and
\citealt{war95} for reviews). As material passes through the inner
Lagrange point of the system towards the white dwarf, the magnetic
field does not initially dominate the (ballistic) motion of the
material. Closer to the white dwarf surface, beyond the stagnation
region, the field threads and disrupts the flow, channelling infalling
material into a funnel which terminates in a shock front at or near
the magnetic pole(s). Shock-heated plasma cools via bremsstrahlung,
Compton cooling, and cyclotron emission as it settles onto the white
dwarf, with the accretion stream also contributing to the optical and
ultraviolet emission. Magnetic interaction between the white dwarf and
its companion keeps the white dwarf in rotational synchronism with the
M~dwarf companion, and the system rotation then leads to the coherent
variability observed in these systems.

The orbital period of UZ~For is 126.5~min, of which the white dwarf is
eclipsed for approximately 8~min (\citealt{oga+88}; \citealt{abs+89};
\citealt{fwb+89}; \citealt{sbt90}; \citealt{bc91}). High time
resolution photometry can probe the structure and dynamics of the
accretion flow, and has been successful in defining the geometry and
emission characteristics of this system (e.g.\ \citealt{abs+89};
\citealt{bc91}; \citealt{is-c98}).

The simultaneous rapid intensity and spectral variations which are
characteristic of the eclipses of cataclysmic variables make these
objects ideal targets for study with advanced photon-counting
detectors recording the time of arrival and the energy of each
incident photon. Although such detectors have long been available for
high-energy studies (e.g.\ proportional counters or CCD detectors
operated in X-ray photon-counting mode), they are only now becoming
available for optical work, based on the new development of
superconducting tunnel junction (STJ) devices (\citealt{pfp93};
\citealt{pvr+96}; \citealt{pvr+97}; \citealt{rpa+98}).

Briefly, a photon incident on an individual STJ breaks a number of the
Cooper pairs responsible for the superconducting state. Since the
energy gap between the ground state and excited state is only a few
meV, each individual photon creates a large number of free electrons,
in proportion to the photon energy. The amount of charge thus produced
is detected and measured, giving an accurate estimate of the photon
arrival time as well as a direct measurement of its energy. Arrays of
such devices provide imaging capabilities.

Following the first laboratory demonstration of the detection
principles (\citealt{pvr+96}), a $6\times6$ array of
$25\times25~\mu$m$^2$ tantalum STJ devices has been built at ESA
(\citealt{rpa+98}; \citealt{rvv+00}). This has been incorporated into
a cryogenic camera operated at the Nasmyth focus of the 4.2-m William
Herschel Telescope on La Palma, where the projected pixel size of
$\sim 0.6\times0.6$~arcsec$^2$ results in an array covering a sky area
of $\sim 4\times4$~arcsec$^2$. Gaps of 4~$\mu$m between array elements
leads to a filling factor of close to~0.8. This camera, `S-Cam2', is a
development of the system first applied to observations of the Crab
pulsar (\citealt{pfp+99}). Several modifications, including a new
detector array, and improved detector stability and uniformity, result
in an improved wavelength resolution of $\Delta\lambda\simeq30$, 60,
and 100~nm at $\lambda=350$, 500, and 650~nm respectively.

Although the intrinsic wavelength response of Ta-based STJs is very
broad, from shortward of 300~nm to longward of 1000~nm, it is
restricted in the present system to about 340--700~nm, as a result of
the atmosphere at the shortest wavelengths and the optical elements
required for the suppression of infrared photons at the longest. The
detector quantum efficiency is around 60--70~per cent across this
range. Instrumental dark current is negligible at the operational
temperature of $0.32\,$K$\sim0.1\,T_{\rm crit}$, and there is no
readout noise.

Detected photons are assigned to energy (or pulse-height analysis,
PHA) channels, formally in the range 0--255, where channels 80--160
cover the most useful response range $\lambda\sim680-340$~nm. Count
rate limits are currently about $5\times10^3$~photons~s$^{-1}$ per
junction, and about $50\times10^3$~photons~s$^{-1}$ over the entire
array. The peak counts for these observations were about 400 detected
photons~s$^{-1}$ on one of the central array elements, just before the
start of eclipse~1.

For each individual detected photon, the arrival time, $x,y$ array
element (or pixel) coordinate, and energy channel are recorded. Photon
arrival times are recorded with an accuracy of about $\pm5$~$\mu$s
with respect to GPS timing signals, which is specified to remain
within 1~$\mu$s of UTC, although typical standard deviations are much
less (\citealt{kus96}). Data are stored as binary FITS files.

The characteristics of STJ arrays are ideally suited to the
observation of CVs.  The high time resolution, high efficiency, large
dynamic range, and modest energy resolution afforded by the S-Cam2
system allow a direct probing of the energy dependence of the
intensity variations across the eclipse, and investigation of the
details of the ingress and egress light curves, whose structure
provides important diagnostics of the emission mechanism. The present
paper presents the first results of a programme of observations of CVs
with this fundamentally new instrument, based on observations of
UZ~For.

\section{Observations and Reductions}
\label{sec:observations}

Observations of three eclipses of UZ~For with the ESA STJ camera at
the William Herschel Telescope were made under reasonable seeing
conditions ($\sim$1~arcsec) during 1999 December: one eclipse on the
night of Dec~9 (with a further one under poorer seeing conditions),
and two on Dec~15 (Table~\ref{tab:observations}).

The S-Cam2 data is of significantly different nature compared to that
produced by CCDs. The data reduction procedure is somewhat more
involved, and is based on the approach normally followed for
high-energy astronomical data. In practice, the off-line analysis
makes extensive use of the {\sevensize FTOOLS} package
(\citealt{bla95x}), developed for the analysis of X-ray data. To obtain
the final data products (i.e.\ energy-resolved light-curves of the
target source, where both energy intervals and time resolution can be
selected), the following steps are applied:

(a) energy calibration: each pixel has a slightly different (although
largely time-independent) energy response, so that photons of the same
energy will be recorded in different energy channels by different
pixels (although the different pixel adjustments are, in practice,
rather small, amounting to only a few PHA channel numbers). The energy
response of each pixel has been initially established using a
laboratory monochromator source, from which the relation between PHA
channel and incident photon energy for each junction has been derived.
The effective temporal stability of the calibration is verified during
the observations by use of an LED source: at the $\sim$10~per cent
energy resolution of S-Cam2 the LED is for all practical purposes a
monochromatic light source. In theory and in practice, the relation
between photon energy and detected charge is highly linear so that,
for example, coincident arrival of two photons of the same energy is
detected as a single photon of twice the energy. When operated at
sufficient intensity, the calibration source therefore yields a
primary peak in the energy spectrum due to the detection of individual
photons, with additional `spectral' peaks at shorter wavelengths due
to the near-simultaneous (sub-$\mu$s) arrival and hence coincident
detection of two and three (and in principle larger numbers of
temporally coincident) photons. From these three peaks the constancy
of the laboratory energy calibration is verified throughout the
observing run, and this information is used to relate the measured
energies (PHA channel numbers) to a common reference. The resulting
`pixel independent' energy calibration yields, for these observations:
\begin{equation}
\lambda({\rm nm})\sim54\,560/(N_{\rm ch}+0.21)
\end{equation}
where $N_{\rm ch}$ is the channel number.

(b) barycentric correction: if required, the photon arrival times can
be corrected to the arrival time at the Solar System barycentre. Our
software uses the JPL DE200 ephemeris, and takes into account Earth
motion, Earth rotation, gravitational propagation delay, and the
transformations from UTC to TAI (+32~s at epoch), from TAI to TDT
(+32.184~s), and from TDT to TDB. Our tabulated eclipse times have
been formally transformed from UTC times of observation to TDB
(barycentric dynamical time). These times should therefore be
decreased by (approximately) $32+32.184\sim64.2$~s for times corrected
only for light travel time effects (Barycentric~JD).

\begin{table}
\caption{\ Observations of UZ~For with S-Cam. $t_1-t_4$ give the times
of the centres of the two sharp drops on ingress and egress
respectively, transformed to TDB (see text), and determined with an
uncertainty of about 0.5~s.}
\vspace{5pt}
\label{tab:observations}
\begin{tabular}{@{}ccccc}
No&	Date&	Obs&	Ingress ($t_1$/$t_2$)&	Egress ($t_4$/$t_3$) \\
&	(1999)&	(UTC)&		(TDB)&	(TDB) \\[4pt]
1&	Dec~09&	21:42--22:52&	22:19:15& 22:27:03 \\
 &	      &		    &	22:19:41& 22:26:25 \\
2&	Dec~09& 02:00--02:41&	\multicolumn{2}{c}{poor seeing} \\
3&	Dec~15&	21:24--22:00&	21:43:01& 21:50:50 \\
 &   	      &	            &   21:43:28& 21:50:12 \\
4&	Dec~15&	23:34--00:18&	23:49:33& 23:57:21 \\
 &	      &	 	    &   23:49:59& 23:56:43 \\
\end{tabular}
\end{table}

\begin{figure}
\epsfig{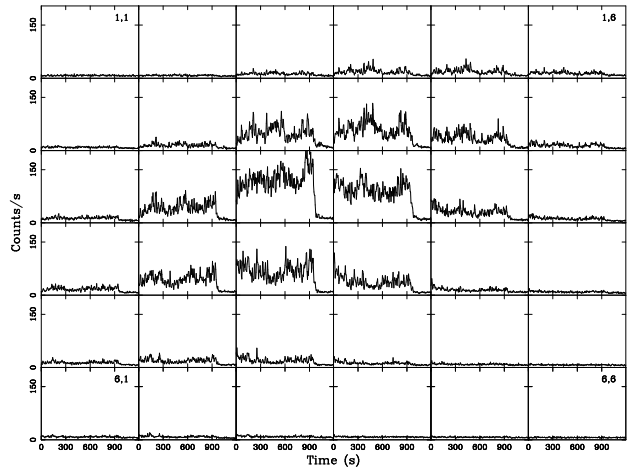}
\caption{Light curve of UZ~For for each of the $6\times6$ array
elements, binned to 5~s, for the 1200~s data interval covering the
ingress of eclipse~1 (second data interval of Figure~\ref{fig:lightcurve})
and for the central wavelength range 490--580~nm. The abscissa (time axis)
covers 0--1200~s, and ordinate (count rate) 0--200~counts~s$^{-1}$.
Note the low and uniform background level at about 7--10~photons
pixel$^{-1}$~s$^{-1}$ for the outer array elements. The corresponding
image for the lower energy interval is displaced marginally towards
the top right, while the image for the higher energy interval is
displaced by almost one pixel to the lower left, shifts due to differential
atmospheric refraction.}
\label{fig:pixels}
\end{figure}

(c) energy range selection: the photon stream is split into a number
of separate files (typically 3--9), according to the energy of each
photon, so that each resulting file represents light of a given
`colour'.  For the results reported here, three energy bands covering
340--490~nm, 490--580~nm, and 580--700~nm were selected. For initial
analysis, this process is automated such that the total counts are
equally distributed amongst the selected energy ranges. Subsequent
analysis steps are applied individually to the resulting files. A
representation of the data at this stage is given in
Figure~\ref{fig:pixels}, where the energy-selected light curves of
each pixel during one of the observations is shown for
one of the selected energy ranges.

(d) correction for the variations in quantum efficiency of the
individual pixels (`flat fielding'): in principle, pixel-to-pixel
sensitivity can depend on both time and energy. Laboratory tests
confirm that the variations in responsivity, $r$, which amount to only
a few per cent pixel-to-pixel ($0.95<r<1.04$, $\sigma_r=0.024$), are
both largely time- and energy-independent. We therefore derive a
single responsivity map from sky observations, and apply it to all
energy ranges. Flat fielding is performed on the data binned into a
specified but arbitrary time interval (e.g., 1~s), facilitating the
subsequent time- or energy-dependent corrections (flat-field response,
extinction correction, and sky background subtraction) which would be
intricate to introduce and deal with at the individual photon level.

In the absence of atmospheric phase fluctuations or telescope-induced
guiding motions, a single flat-field correction could be applied to the
whole observation (`exposure-time correction' in X-ray astronomy).
However, any resulting image motion on the array implies that
uncorrected spatial variations in response would contribute apparent
fluctuations in the object's light curve as a function of time. Our
adopted procedure allows time-independent response corrections to be
applied in the presence of such time-dependent image motion.

(e) correction for atmospheric extinction: the time-binned and
flat-fielded data are adjusted for extinction using the standard ING
(Isaac Newton Group) tabulation of extinction values as a function of
wavelength and air mass.

\begin{figure}
\epsfig{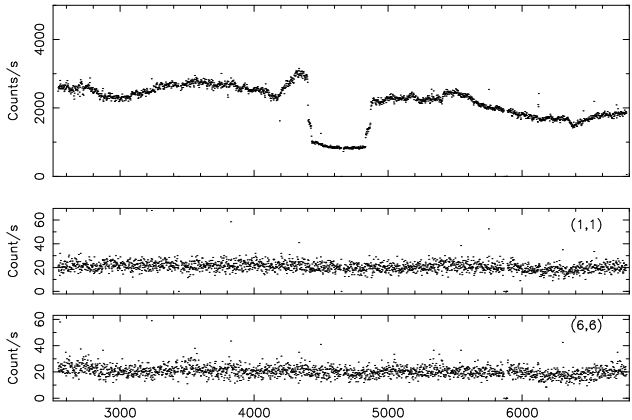}
\caption{The light curve of UZ~For centred on eclipse~1. The data acquired 
over 70~min comprised four exposures (data files) of 900+1200+1200+900~s
(small resulting gaps of a few seconds in the data coverage are
evident). The data, acquired with a time resolution of 5~$\mu$s, has
been binned to 1~s, and is the sum of the total array signal (over
36~junctions) without background subtraction. The bottom two panels
show light curves of array elements (1,1) and (6,6) at the array
edges, used for sky background determination and showing their
stability. The eclipse minimum lies slightly, but significantly, above
the measured sky background (Figure~\ref{fig:colour1}). The data
covers an interval of about $\pm0.28$ in orbital phase, centred on the
eclipse.}
\label{fig:lightcurve}
\end{figure}

(f) sky background subtraction: this is complicated by the small
size of the array, and by the undersampling of the time-varying seeing
profile. Our current approach is non-optimal but adequate for
present purposes, and is performed on the time-binned data constructed
in step~(d).  From a visual inspection of the pixel-to-pixel
light curves as a function of energy (e.g.\ Figure~\ref{fig:pixels})
we can define a mask comprising `source' and `background' (as well 
as possibly `rejected') array elements. In the present analysis all
pixels could be used. The source intensity was therefore simply
derived from the sum of all array elements, and the sky contribution
derived from two corner pixels (Figure~\ref{fig:lightcurve}). This
yields a set of background-corrected light curves, in different
energies, at the selected binning period. A more optimum
source/background mask will depend slightly on energy, due to the
effects of differential atmospheric refraction, which can cause a
displacement of the (energy-dependent) image centroid by up to 1~array
element (0.6~arcsec) for reasonably large zenith angles.  In practice,
as seen in Figure~\ref{fig:pixels}, corner pixels (1,1) and (6,6)
provide a robust estimate of the sky background. This is also shown in
Figure~\ref{fig:lightcurve}, where the raw (uncalibrated) light curve
for eclipse~1 is shown, for the whole array and for the two corner
pixels.

\begin{figure}
\centerline{\epsfig{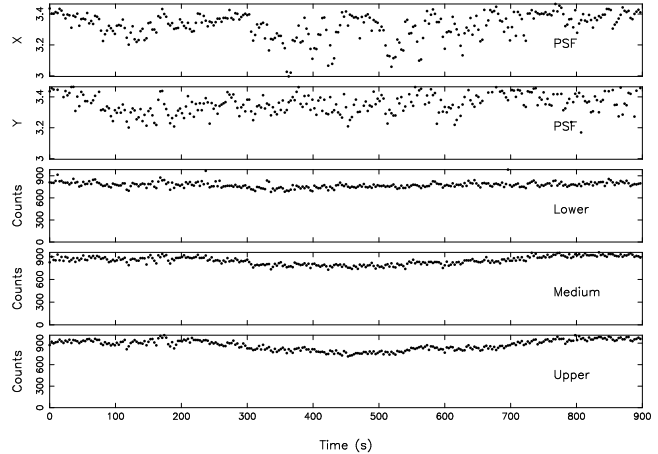}}
\caption{The image centroid positions in $x$ and $y$ (in pixels), 
determined over intervals of 3~s (top two panels), compared with the 
corresponding intensity variations of the source summed over the 
array for the three energy ranges (lower three panels). The data 
are from the first data interval of eclipse~1 (cf.\ the lower 
intensity seen at around 3000~s in Figure~\ref{fig:lightcurve}). The
absence of prominent correlations between position and intensity
provides further confirmation of the data quality.}
\label{fig:motion}
\end{figure}

As described under step~(d), if the energy-dependent pixel-to-pixel
response variations are large (and uncorrected), atmospheric-induced image 
motion would affect the light curves through the interdependence of image
position and array sensitivity. We have performed experiments in which
the data are binned into, e.g., 1~s time bins, and the image centroid
positions in $x$ and $y$ are compared with the corresponding intensity
variations (Figure~\ref{fig:motion}). In practice, the image centroid
moves by at most $\pm0.1$~pixels, and even for our data before energy
calibration there is no evident correlation between image motion and
intensity flickering.

A small part of the high-frequency structure of the light curves can
formally be attributed to the time-dependent loss of flux into the
small inter-pixel gaps (dead zones) between the array elements,
as the atmospheric phase fluctuations move the `seeing profile' around
the array. Some improvement in this instrumental contribution to the
light curve flickering could be made by taking into account the
varying contribution to the dead zones as a function of image
location. This could be done by determining a Gaussian profile fit to
the data binned over, say, 1~s, taking account of the dead zones in
the fitting process, then normalising the measured flux to the total
area covered by the image during that interval. The effect is
expected to be rather small, and a detailed study is deferred to 
a future paper.

The pixel-to-pixel light curves (e.g.\ Figure~\ref{fig:pixels}) show 
structure inconsistent with Poisson noise, and only roughly correlated 
between neighbouring pixels. From the clean behaviour of the total
light curve, we infer that this structure is consistent with
atmospheric phase fluctuations, and again return to this issue in a
future study.

To assess the fidelity of our data further, we have investigated some
specific features of the light curves in detail. For example, one low
data point appears as a single outlier in all three energy bins at
about phase 0.938 in the 2~s binned data of Figure~\ref{fig:colour1}.
At a time binning of 0.02~s, this feature is seen to arise from a
rapid but not instantaneous drop to zero source counts, which persists
for almost 1~s, and which can be attributed to a known but 
infrequent telescope pointing glitch in azimuth, with an amplitude of
several arcsec (C.R.~Benn, private communication).

\begin{figure*}
\centerline{\epsfig{file=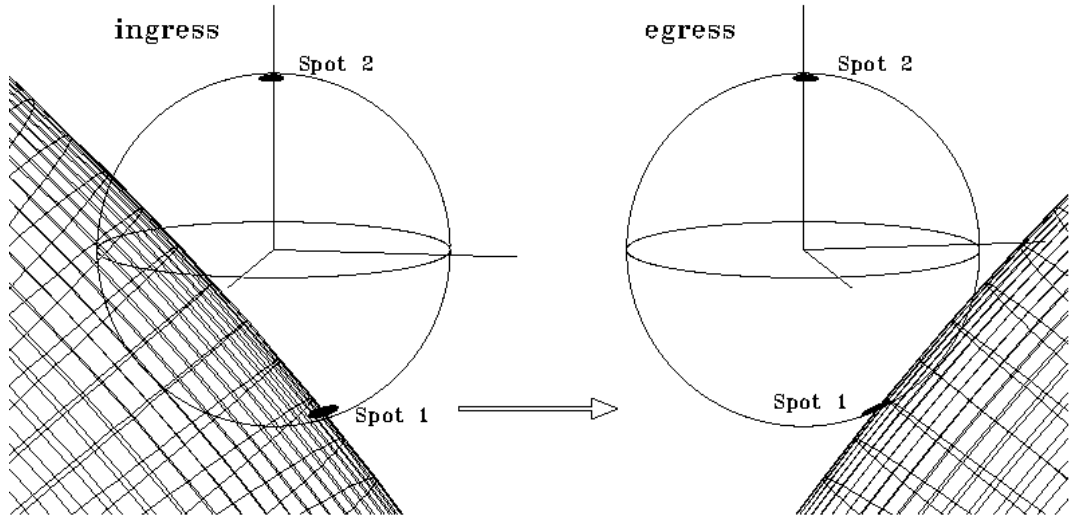,height=15.0cm,angle=270}}
\caption{View of the system at phases of ingress and egress of spot~1 for
the parameters of eclipse~1 for $q=0.2$ in Table~\ref{tab:regions}. The
secondary Roche lobe is projected as a wire model so that the white dwarf
remains visible. The system inclination, $i$, where $i=0^\circ$ 
corresponds to pole-on, is inferred to be about $80^\circ$. The 
origin of the coordinate system is at the centre of the white dwarf, 
with the origin of co-latitude vertically `up' (so that spot~2 has a 
co-latitude of close to $0^\circ$, while spot~1 is around $150^\circ$). 
The origin of longitude is the coordinate axis joining the line of
centres, with increasing longitude in the direction of the third
coordinate triad. Ingress or egress of a given spot constrains its
instantaneous projected location on the plane of the sky to lie along
the limb of the secondary at that instant. The combination of timings
of both the covering and uncovering of a given spot then constrain its
position in two dimensions.}
\label{fig:spotview}
\end{figure*}

\section{Results and Discussion}
\label{sec:results}

\subsection{Geometrical overview}
\label{sec:geometry}

Figure~\ref{fig:lightcurve} displays a number of prominent features
observed previously in UZ~For through the optical (e.g.\
\citealt{abs+89}; \citealt{bc91}; \citealt{is-c98}) to ultraviolet
(\citealt{wsv95}; \citealt{ss96}), and related to the variable viewing
geometry with orbital phase (cf.\ Figure~8 of \citealt{fwb+89};
Figure~4 of \citealt{bc91}; Figure~5 of \citealt{wsv95}; Figure~7 of
\citealt{ss96}). Most notable are the intensity variations out of
eclipse and especially the pre-eclipse brightening, and the rapid
changes during eclipse ingress and egress, attributable to the
variable viewing geometry of the accretion spots and the accretion
stream (e.g.\ \citealt{hch+99}; \citealt{kgb00}).

Our present understanding of this system is illustrated in
Figure~\ref{fig:spotview}, where the locations of the two accretion
spots resulting from the present investigation are illustrated. The
white dwarf is rotationally locked to the secondary, and eclipsed for
about 8~min as the system rotates. Covering/uncovering of the white
dwarf photosphere during ingress/egress occurs in about 40~s. The
projected positions of the accretion spots on the plane of the sky are
fully defined by the times of the successive rapid changes in light
intensity as the spots themselves are covered and uncovered. If the
position of the accretion spots are assumed to be located at the white
dwarf surface, their location on the white dwarf is then defined by
the inferred white dwarf radius. The rate of accretion affects the
system brightness, and only at low accretion rates (i.e.\ in the low
state) can emission from the white dwarf photosphere be distinguished
(\citealt{bc91}). The location of the emission regions probably
depends on accretion rate: even it is assumed that the spin and
magnetic field axes are aligned, the accretion spots can still be
located away from the poles (e.g.\ towards the secondary) depending on
where the ballistic flow impacts the field lines. With variable higher
accretion rates, the accretion spots may therefore be expected to move
accordingly.

\subsection{System brightness}

The optical light curve of UZ~For is dominated by accretion flow onto
the primary, and changes significantly from epoch to epoch.
\citet{is-c98} identify three states, `low' (V$\sim18.5$), `faint
high' (V$\sim 16.7$) and `bright high' (V$\sim 15.9$), and provide a
schematic representation of these. Our estimate of the magnitude of
UZ~For at the epoch of observation is derived from the observation of
a $V=7.1$~mag star through a ND4 neutral density filter, resulting in
3300 detected photons$^{-1}$ over the array. With about
700~photons$^{-1}$ over the whole array from the sky, the
sky-corrected signal in Figure~\ref{fig:lightcurve} corresponds to a
magnitude of about 17.7--18.1~mag out of eclipse, depending on phase,
and ignoring the different spectral energy distributions of UZ~For and
the comparison star. At the time of our observations, the system
therefore appears to be in an `intermediate' state between the low
state of \citet{bc91} and faint high state of \citet{is-c98}.

In the low state, the light curve has an on-off appearance typical of
a system accreting at a single point on the white dwarf, corresponding
to the location of spot~1 in Figure~\ref{fig:spotview}. When this is
not in view, the majority of the emission from the system at phases
between $\phi\sim0.15-0.65$ is from the two stellar photospheres. In
the higher states, this phase interval is increasingly bright, until
in the bright high state, the system is brightest at $\phi\sim0.25$.

\subsection{Shapes of eclipses and eclipse timing}
\label{sec:eclipses}

The eclipse morphology of our data is distinct from that in the low
and high states. Two pairs of rapid changes in brightness are
prominent in all of our eclipses, around phases 0.97 and 1.03
(Figures~\ref{fig:colour1} and~\ref{fig:eclipses}), in contrast with
the single active accretion pole observed by \citet{is-c98}. These can
be assigned to the successive covering and uncovering by the secondary
of two small accretion regions (`spots') on the surface of the white
dwarf. These regions are resolved in our data at 0.5~s time
resolution, with ingresses and egresses lasting $\sim$1.5~s
(Figure~\ref{fig:eclipses}). Such behaviour is also visible (but not
discussed) in \citet{bai95}, when the system is $\sim$1~mag brighter
than the low state, and thus also in the intermediate state. The
observations of \citet{fwb+89} were also in the intermediate state,
but the time resolution of the data is insufficient to resolve rapid
intensity changes through eclipse.

\citet{bai95} noted that the outer pair of rapid changes (spot~1,
Figure~\ref{fig:colour1}) coincide in phase with the low state
eclipse, and also with the simultaneous soft X-ray eclipse seen using
{\it ROSAT}. This pair also coincide with the rapid changes in
brightness in the faint high and bright high state data in
\citet{is-c98}, the eclipses seen in the EUV by \citet{wsv95}, and
those in the UV by \citet{ss96}. Spot~2, in contrast, is not known to
emit in X-rays.

\begin{figure}
\begin{center}
\epsfig{file=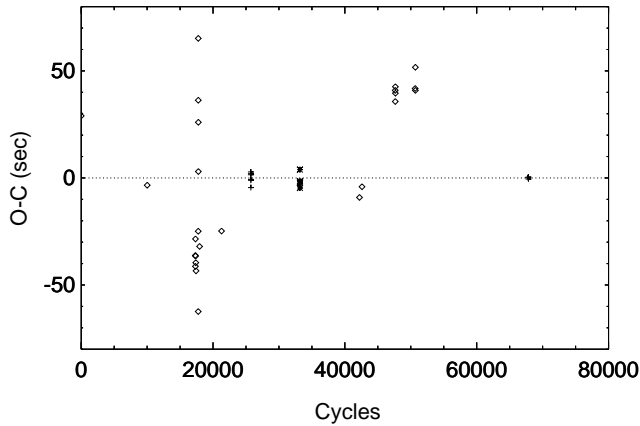,angle=90,width=9.2cm,
	bbllx=0,bblly=320,bburx=310,bbury=800,clip=}
\end{center}
\vspace{-15pt} 
\caption{Timing residuals with respect to our revised ephemeris 
(Equation~\ref{equ:ephem}) constructed from the eclipse timings 
in \protect\citet{bc91} and our present data (crosses). Also shown for
comparison are the {\it ROSAT\/} timings from \protect\citet{ram94} (stars),
and other timings in the literature (diamonds).}
\label{fig:timings}
\end{figure}

The shapes of the three eclipses in Figure~\ref{fig:eclipses} is
similar. However, there are clear differences in the amplitude of the
rapid brightness changes from eclipse to eclipse. Both spots are
fainter in eclipse~3 compared to eclipse~1 and~4. There are also
differences in the relative brightnesses at ingress and egress: the
brightness change at ingress is larger for eclipse~1, while that at
egress is larger for eclipse~4. In general however the relative spot
total brightnesses are constant, with spot~1 $\sim$3 times brighter
than spot~2. This indicates to first order that the relative accretion
flow to each spot remains constant even in the presence of changes in
the total flow rate.

In Table~\ref{tab:observations}, we estimate the mid-points of each of
these pairs of rapid changes in brightness for three of the four
eclipses observed. For each eclipse, the outer drops
($t_{1}$~and~$t_{4}$) are the most precipitous. There is no evidence
for differences in these times as a function of energy, and our
estimates of the mid-times are probably accurate to about 0.5~s.  As
noted in Section~\ref{sec:observations}, our times have been formally
transformed from UTC times of observation to TDB (barycentric
dynamical time).

The eclipse ingress and egress durations for the spots in the three
eclipses in Figure~\ref{fig:eclipses} are the same for each eclipse to
within the measurement uncertainties. There is some evidence that the
ingress duration for spot~1 is longer than the egress ($\sim$2~s as
opposed to $\sim$1.5~s). These durations are similar to those in
\citet{bc91} (but where the time resolution was barely sufficient).
The ingress and egress durations for spot~2 are only slightly less
than those for spot~1 and of equal duration in the three eclipses,
indicating a spot radius similar to that of spot~1. These durations
are also consistent with upper limits in the EUV (\citealt{wsv95}) and
soft X-ray (\citealt{bai95}), of about 4.7~s and $\sim$5~s
respectively.

Only in the low state of \citet{bc91} is the white dwarf clearly
detectable, and hence usable as a timing reference. Most UZ~For studies
(\citealt{wsv95}; \citealt{is-c98}) therefore use the spot~1
mid-eclipse point, corresponding here to the outer pair,
$(t_1+t_4)/2$, corrected to the centre of the white dwarf eclipse (by
subtracting 9~s as appropriate for the $45^{\circ}$ longitude in
\citealt{bc91}), as the origin of zero phase. Eclipse~1 represents
eclipse $67\,777$ (about 16~years) after the origin given by
\citet{is-c98}. The cycle count is unambiguous, but our eclipses occur
$\sim$100~s earlier than predicted by their quadratic ephemeris.

The eclipse timings in the literature (\citealt{ram94};
\citealt{wsv95}; \citealt{is-c98}) show timing residuals of $\pm$50~s
from the \citet{is-c98} ephemeris. Part of the timing residuals may be
attributable to motion of the spot on the surface of the white dwarf,
for example as a result of changes in the accretion rate. However,
these changes cannot amount to more than the eclipse duration of the
white dwarf ($\sim$40~s), unless the emission originates far above the
surface, and it seems unlikely that the more extreme residuals can be
accommodated in this way. In view of the heterogeneous data, the large
residuals with respect to the \citet{is-c98} ephemeris, and the
uncertainties in correcting to the centre of the white dwarf, we have
derived a linear fit using only our high time resolution data (three
eclipses) and that with the next best time resolution (from
\citealt{bc91}, six eclipses) to define a linear ephemeris:
\begin{equation}
\label{equ:ephem}
{\rm BJD} = 2\,445\,567.17653(2) +0.0878654325(4)\, {\rm E} 
\end{equation}
where BJD is the barycentric Julian Date, E~is the eclipse number and
the uncertainties in brackets are in the last figure. The resulting
fit (Figure~\ref{fig:timings}) leads to residuals of less than 1~s for
our data, less than about 4~s for the \citealt{bc91} data, and less
than about 5~s for the {\it ROSAT\/} timings of \citet{ram94}. This
suggests that the large residuals noted previously may result from
timing errors (or untraceable time-scale inconsistencies) in these
different datasets. High precision timing data will be important for
future investigations of the spot longitude changes and correlations
with accretion rate.

\subsection{Geometrical interpretation}

Our precise timing and high-speed photometric data allow the position
of the accretion spots to be located on the plane of the sky
(Figure~\ref{fig:spotview}). To locate them on the white dwarf,
assuming that they originate at the surface, we require an estimate of
the white dwarf radius. Working in terms of the system mass ratio,
$q$, we trace the Roche potential out of the binary system along the
line of sight from any point in the vicinity of the white dwarf. We
then adjust the system parameters under the constraints of grazing
contact of this line of sight with the Roche lobe at particular
phases. Starting with the generally favoured $q=0.2$ solution of
\citet{bc91} to calculate the positions of the accretion regions on
the white dwarf (corresponding roughly to $R_{\rm
WD}\sim8\times10^8$~cm, and $M_{\rm WD}\sim0.7\,M_\odot$, cf.\
\citealt{smb97}), we recover the radius of the white dwarf and
co-latitude of the spot seen in their data and given in their Table~2,
to the level of accuracy recorded. The derived parameters are
sensitive to small changes in inclination and white dwarf radius, and
this is sufficiently accurate to ensure a common basis for the
following comparisons.

\begin{table}
\caption{\ Inclination and co-latitudes/longitudes of spots~1 and~2 as
a function of mass ratio~$q$, assuming both spots are in the hemisphere
facing the secondary (however, see Section~\ref{sec:brightness}).
Figure~\ref{fig:spotview} should be consulted for coordinate
definitions. Errors are estimated at the $1\sigma$ level.}
\vspace{5pt}
\label{tab:regions}
\begin{tabular}{@{}llcccr}
Eclipse & Inclination & \multicolumn{2}{c}{Spot~1} & \multicolumn{2}{c}{Spot~2} \\
number &   & Co-lat & Long & Co-lat &	Long \\
 & ($^{\circ}$)& ($^{\circ}$)& ($^{\circ}$)& ($^{\circ}$)& ($^{\circ}$) \\[4pt]
\multicolumn{6}{l}{$q=0.15$}   \\
1 & $82.7\pm0.1$ & $144\pm3$ & $31$ & $0\pm3$ & $-3\pm10$ \\ 
3 & $82.7\pm0.1$ & $149\pm3$ & $31$ & $0\pm3$ & $-3\pm10$ \\ 
4 & $82.7\pm0.1$ & $144\pm3$ & $31$ & $0\pm3$ & $-3\pm10$ \\ 
\multicolumn{6}{l}{$q=0.2$}   \\
1 & $80.9\pm0.1$ & $150\pm3$ & $45$ & $4\pm3$ & $-10\pm7$ \\ 
3 & $80.9\pm0.1$ & $148\pm3$ & $45$ & $4\pm3$ & $-10\pm7$ \\ 
4 & $80.9\pm0.1$ & $150\pm3$ & $45$ & $4\pm3$ & $-10\pm7$ \\ 
\multicolumn{6}{l}{$q=0.3$}   \\
1 & $78.5\pm0.1$ & $139\pm3$ & $45$ & $11\pm3$ & $-10\pm8$ \\ 
3 & $78.5\pm0.1$ & $144\pm3$ & $45$ & $11\pm3$ & $-10\pm8$ \\ 
4 & $78.5\pm0.1$ & $136\pm3$ & $45$ & $11\pm3$ & $-10\pm8$ \\ 
\end{tabular}
\end{table}

We have measured the eclipse durations for spot~1 in our data, in the
{\it EUVE\/} data of \citet{wsv95}, and in the {\it ROSAT\/} and
optical data of \citet{bai95}. As in \citet{bc91} and other previous
studies, we assume the accretion spots are located on the surface of
the white dwarf. If we allow the phase to be a free parameter to
accommodate the timing residuals against the ephemeris (see
Section~\ref{sec:eclipses}), spot~1 can be located at a co-latitude of
$\sim$150$^{\circ}$ and longitude of $\sim$45$^{\circ}$ (ahead of the
line of centres) from all these data, as in \citet{bc91}. The view at
the phases of ingress and egress of spot~1 is shown in
Figure~\ref{fig:spotview}. There is no evidence for any movement of
this region in co-latitude ($<5^{\circ}$). Any longitude changes from
epoch to epoch are mixed in with the timing uncertainties, so these
are not possible to determine.

From our data for $q=0.2$ we locate spot~2 in the `upper' hemisphere
(that inclined towards the line of sight), at a co-latitude of
$4^{\circ}\pm3^{\circ}$ and a longitude of $-10^{\circ}\pm7^{\circ}$
(Figure~\ref{fig:spotview}). A similar fit to the \citet{bai95} data
yields $5^{\circ}\pm3^{\circ}$ and a longitude of
$-10^{\circ}\pm12^{\circ}$. Spot~2 is therefore close to the rotation
axis and slightly behind the line of centres, and at the same
co-latitude at these two epochs.

UZ~For provides the opportunity to obtain highly constrained
fundamental system parameters in an interacting binary.  We summarise
the locations of the spots in Table~\ref{tab:regions} for $q=0.2$ and
the other mass ratios in the range $0.15-0.3$ found to be acceptable
by \citet{bc91}. For $q<0.15$ no solution is achievable for spot~2
unless it is above the surface of the white dwarf. In order to
progress further with $q$, we have to update the mass-radius
relationships used by \citet{bc91}. Recent measurements of white dwarf
masses and radii for 40~Eri~B (\citealt{sph+97}) and V471~Tau
(\citealt{bhc+97}) indicate close agreement with the Hamada-Salpeter
mass-radius relation for He white dwarfs used by \citet{bc91}. Recent
mass-radius data derived using Hipparcos parallaxes (\citealt{vsk+97})
do not add to the accuracy. No progress is therefore possible in the
case of the primary. With respect to the secondary, both the
period-mass relations of \citet{war95} and \citet{pat84} predict a
secondary mass $M_2=0.17\,M_\odot$, consistent with $q=0.23$, while
the \citet{kb00} ZAMS models predict a secondary mass of
$M_2=0.24\,M_\odot$, consistent with $q=0.3$. Unless the secondary was
significantly evolved at the onset of mass transfer (\citealt{kb00}),
values of $q=0.2-0.3$ are to be preferred, implying $M_{\rm
WD}=0.7-0.8\,M_\odot$. This would require the better fits to the white
dwarf eclipse data in \citet{bc91} at $q=0.15$ than $q=0.2$ to result
from their neglect of limb darkening. This conclusion rests, however,
on the reliability of the secondary period-mass relationship, which
has been the subject of considerable discussion (see \citealt{war95}).

These locations are consistent with the spot eclipse ingress and egress
durations (Section~\ref{sec:eclipses}): as is evident from Figure
\ref{fig:spotview} the projected dimensions of spot~1 are smaller in
egress, leading to a shorter egress than ingress duration, as
observed. On the other hand the projected dimensions of spot~2 are
almost unchanged between ingress and egress, suggesting similar
ingress and egress durations, again as observed. From the duration of
the spot ingress and egress, we calculate that the angle subtended by
either spot from the centre of the white dwarf in the dimension of the
secondary limb travel is $3^{\circ}\pm1^{\circ}$, in all eclipses. For
a circular spot this translates to a fractional area of
$7\times10^{-4}$, consistent with the upper limit of $5\times10^{-3}$
in \citet{bc91}.

Further progress in determining the parameters for this system is
desirable and feasible by obtaining the highest quality measurements
of the eclipse profile in the low state.

\begin{figure*}
\begin{center}
\epsfig{file=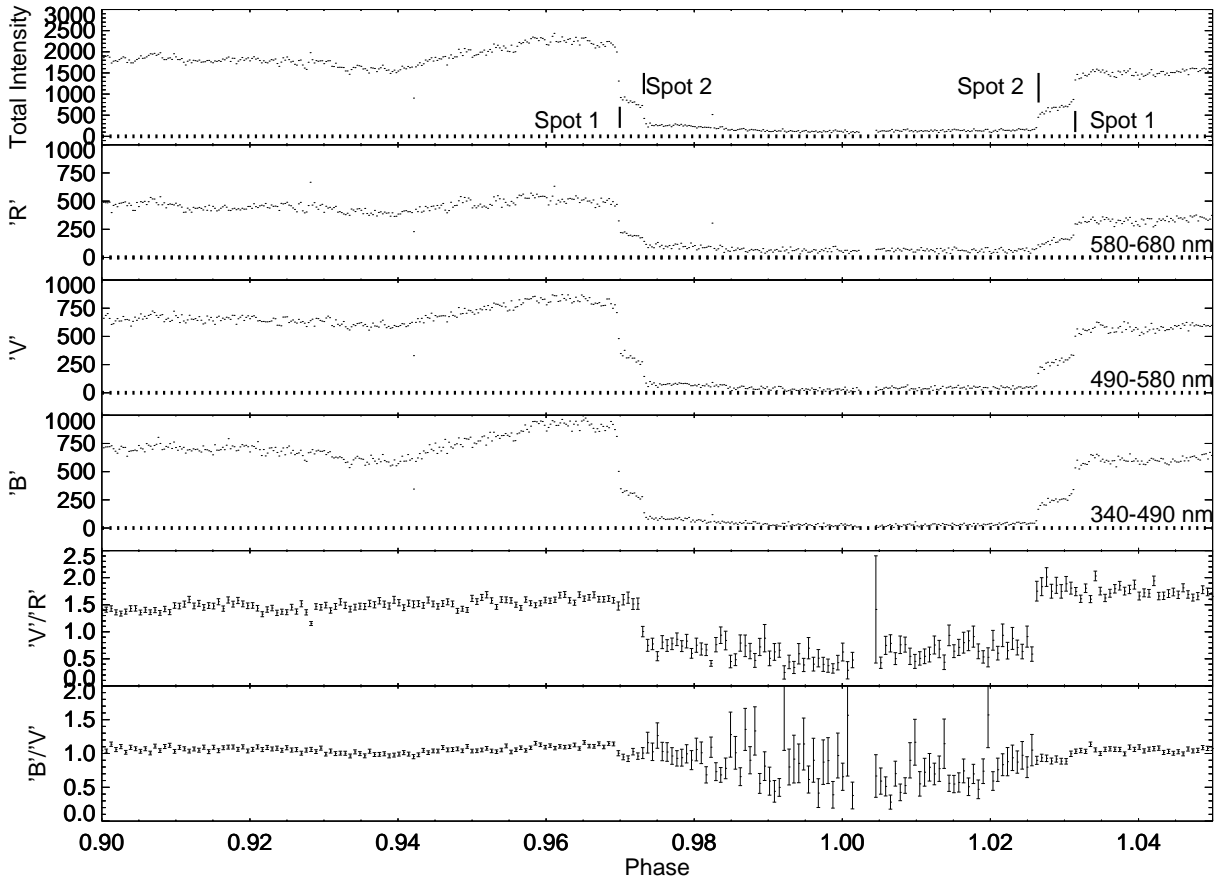,width=12cm,angle=90}
\end{center}
\caption{Sky-subtracted and flat-fielded total counts s$^{-1}$ (top),
counts s$^{-1}$ in three bands 340--490 nm (labelled `B' for ease of
reference), 490--580 nm (`V') and 580--700 nm (`R'), and two colour
ratios constructed from the three energy-resolved light curves for
eclipse~1. Lower values of these ratios imply a redder colour. Data
are displayed in bins of 2~s for the intensities, and 5~s for the 
colour ratios, with error bars for the ratios corresponding to
photon statistical errors. Orbital phase is with respect to the
updated ephemeris given by Equation~\ref{equ:ephem}. }
\label{fig:colour1}
\end{figure*}

\subsection{Colour variations}

\begin{figure}
\centerline{\epsfig{file=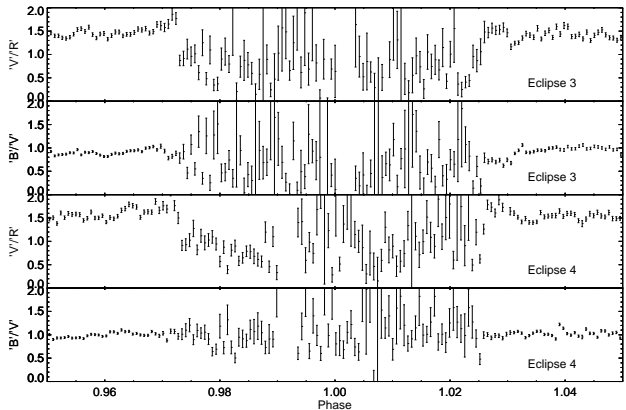,angle=90,width=8.5cm}}
\caption{Sky-subtracted and flat-fielded colour ratios for eclipses 
3~and~4, binned at 5~s (as for Figure~\ref{fig:colour1}).}
\label{fig:colour2}
\end{figure}

\begin{figure}
\epsfig{file=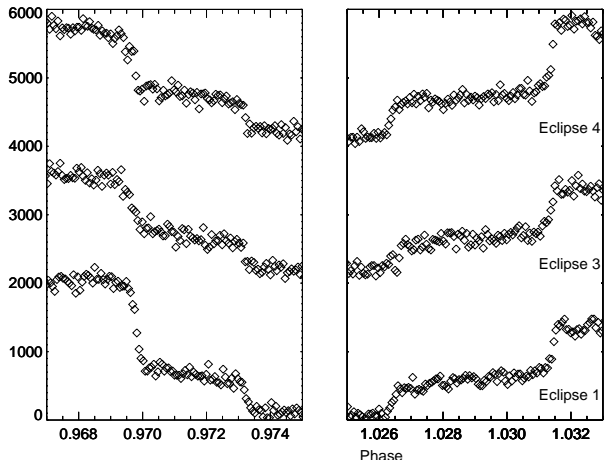,width=8.5cm}
\caption{The white light ingresses and egresses of eclipses 1, 3 and 4
at 0.5~s time resolution (see Table~\ref{tab:observations}). Each
successive eclipse is displaced vertically by 2000 counts s$^{-1}$. 
Orbital phase is with respect to the updated ephemeris given by
Equation~\ref{equ:ephem}.}
\label{fig:eclipses}
\end{figure}

Figure~\ref{fig:colour1} provides the background-subtracted
energy-resolved light curves for eclipse~1, and also the two colour
(hardness) ratios. For conciseness we refer to these as V:R and B:V
ratios (from `red', `visual' and `blue' based on the bandpasses in
Figure~\ref{fig:colour1}). We also show the colour ratio changes in
eclipses~3 and~4 in Figure~\ref{fig:colour2}. Eclipses~1 and~3 are
similar, while the B:V ratio in eclipse~4 shows less variation than in
the first two.

After both spots are eclipsed, the colour is significantly redder in
the V:R ratio, in line with our expectation that the major contributor
to the `red' band at these phases is the secondary star. Although also
redder at mid-eclipse, the B:V ratio has a somewhat different
behaviour, generally decreasing to mid-eclipse and then increasing
again. No significant change in colour is expected from the secondary
during the narrow phase range spanned by the eclipse. This indicates
that flux contributed by the secondary in these bands is less
significant, as expected for a dwarf M~star.  The white dwarf and
associated accretion regions on its surface will have been covered
within $\sim$40~s of eclipse spot ingress (see \citealt{bc91}), so by
mid-eclipse the flux contributing to the colour change must be some
distance from the white dwarf. The only obvious candidate is the
accretion flow between the two stars. However, the approximately
symmetrical B:V ratio change though eclipse is not easily reconciled
with emission from a ballistic+magnetic trajectory to only spot~1. The
stream to spot~2 is uncovered before spot~2 egress and when the stream
to spot~1 is still eclipsed. It is therefore likely that the B:V
colour variation is due to a combination of the progressive covering
and uncovering of both streams.

Care is required in the interpretation of the colour changes during
eclipse, as the background contribution (from, for example, the
secondary star) affects the interpretation of the colour ratio
changes. Nevertheless the colour ratios in Figure~\ref{fig:colour1}
provide evidence for significant changes at the time of the spot
ingresses and egresses, in the sense that the V:R ratio becomes
marginally bluer and the B:V ratio somewhat redder once spot~1 is
eclipsed. This behaviour is repeated in eclipses~3 and~4 (although not
in the B:V ratio for eclipse~4). It is possible to separate out with
some confidence the contributions of each of the spots, stream and
secondary in the three bands for the three eclipses.  This indicates
that the relative contribution of spot~2 in comparison to spot~1 is
$0.52\pm0.14$, $0.58\pm0.10$ and $0.36\pm0.03$ in the B, V and R bands
at ingress and $0.54\pm0.07$, $0.71\pm0.16$ and $0.38\pm0.15$ at
egress respectively. Although there appear to be small differences in
the colours from eclipse to eclipse, the statistical evidence for this
is weak.

The relatively greater contribution in V may indicate that there are
cyclotron humps in the spectrum of spot~2, not a broad continuum, and
if so, is indicative of lower temperatures.  Such humps have been seen
out of eclipse by \citet{fwb+89} and \citet{sbt90}, who found a lower
magnetic field in spot~2 compared to spot~1, but it is unclear whether
these can explain the relative flux distributions. An alternative
explanation may lie in the different viewing angles to the two spots
at the time of eclipse: spot~2 is seen almost face on, with bluer
harmonics undergoing more absorption and scattering at these viewing
angles (\citealt{wm85}), while spot~1 is seen almost side on.

There is also a slow general colour variation in the V:R ratio over the
duration of the observations, with the source tending to become bluer
towards the time of eclipse, although the colour changes more
substantially during the eclipse itself. We can investigate whether
the slow variation might originate from the ellipsoidal variation of
the red light from the secondary (\citealt{fwb+89}).  An upper limit for
the emission from the secondary is available from the count rate near
mid-eclipse, $\sim$4~per cent in~R, which is therefore insufficient to
cause the observed V:R ratio change. This suggests that the V:R change
must be due to a difference in cyclotron flux distributions from the
spots, either because of intrinsic differences, or because of different
viewing angles to the spots at any one phase.

In this context it is interesting to note that when spot~1 disappears
over the limb at $\phi\sim1.15$ ($\sim$5780~s in
Figure~\ref{fig:lightcurve}) there is no significant colour change in
V:R or B:V.  This indicates that when viewing angles are similar, the
apparent flux distributions are also similar, and so viewing angle
differences to the two spots are likely to be the main cause of the
slow variation in V:R outside of eclipse. Spot~2 is seen through a
greater range of angles than spot~1, so this explanation is consistent
with the fact that the V:R variation is symmetric about a point near
the end of the eclipse, which is when spot~2 is seen most face on
(Table~\ref{tab:regions}).

There is a reduction in brightness before the eclipse at
$\phi\sim0.93$ (Figure~\ref{fig:colour1}, and $\sim$4200~s in
Figure~\ref{fig:lightcurve}). Although from these optical data it is
could be argued that the pre-eclipse dip is a photometric variation
(see also \citealt{bc91}), it is clear from EUV data (\citealt{wsv95})
that the dip is caused by absorption in the stream crossing the line of
sight to the accretion regions. The dip is almost grey, but slightly
stronger in the blue. This is somewhat different from that seen in
HU~Aqr by \citet{hch+99}, where the dip is strongest at red
wavelengths. The almost grey colour of the dip indicates that the
extinction must be caused mostly by electron scattering or by
completely obscuring material, although the latter is unlikely
throughout the stream cross section.

\subsection{A search for high frequency variations in the light curves}

Several AM~Her systems have been observed to show quasi-periodic
oscillations (QPOs) on time-scales of 1--2~s (\citealt{mid82};
\citealt{lar87}). Their origin in the shock region was conclusively
demonstrated by \citet{lar89} who found that QPOs were not observed
when the shock region disappeared from view as it rotated behind the
white dwarf. No evidence for QPOs has yet been found in UZ~For
(\citealt{is-c98}). Our data allow us to search for evidence of QPOs
with significantly higher sensitivity.

The data were split into several sections and the three energy bands
defined, as previously noted, so that an approximately equal number of
counts were present in each. A time binning of 0.2~s was used for each
band. A Discrete Fourier Transform (DFT) was used to search for
periodic signals in the data. No evidence for a QPO in any energy band
was detected. We can set an upper limit for a strictly periodic
signal of $\sim$0.5~per cent in each band. Since the power of a QPO is
spread in frequency and not strictly periodic, it is more difficult to
give an accurate upper limit for a QPO, but we expect this also to be
$\sim$0.5~per cent.

The frequency of the shock oscillations depends sensitively on the local
cooling rate within the shock, which varies with $\dot{M}$ and the
magnetic field strength. In magnetic systems there are two competing
cooling mechanisms in the accretion shock: thermal bremsstrahlung and
cyclotron radiation. \citet{cls85} showed that when cyclotron cooling
was strong, the shock tended to be stabilised against oscillations.
Further, \citet{wpc+96} found that the stabilising influence depends
on the magnetic field strength: the higher the field strength the
greater the stability. This was explored further by \citet{sw99}. In
the case of UZ~For, \citet{sbt90} found that the main accretion pole
had a field strength of 53~MG and the secondary pole a field strength
of 75~MG. This would suggest that cyclotron emission is the dominant
cooling process in each accretion region and the shock is relatively
stable -- as confirmed by these observations.

\subsection{Behaviour in different brightness states}
\label{sec:brightness}

We now consider the origin of the different photometric behaviour at
different brightness states of the system.

The low state maximum is $\sim$0.1 of the bright high state maximum,
indicating a substantial increase in emission in the higher states,
particularly at those phases $\sim$0.15--0.65 when spot~1 is out of
view. Spot~2 is in view between $\sim$0.75--0.35, with maxima expected
from cyclotron beaming (\citealt{wm85}) at phases 0.85 and 0.25, or at
phase 0.05, depending on the spot co-latitude. Some of the additional
contribution in the bright high state could therefore come from this
spot, but not the majority because spot~2 is at most only weakly
evident in the eclipse profiles in Figures~7 and~8 of \citet{is-c98}.

These considerations, and the phasing of the slow increase after
spot~1 egress in the bright high state data, can only arise, as argued
by \citet{is-c98}, from an extended emission structure leading the
line of centres -- probably the accretion stream to spot~1. At
ingress, spot~1 is itself $\sim$5 times brighter than it was in the
low state ($\sim$2 times at egress). \citet{is-c98} note that the
accretion stream is at least as bright as spot~1 at ingress, and
significantly brighter than any contribution from spot~2 that might be
buried in their light curves. Given that spot~1 will have passed out
of view over the limb at phase 0.15, and given the absence of strong
emission from spot~2, almost all of the emission in the bright high
state at phase~0.25 (at which phase the system is at its brightest)
must be from the stream.

The rotation of spot~1 over the limb should cause a change in
brightness. None of our observations covers the phases when the spot
rotates into view, but the rotation out of view is covered in
Figure~\ref{fig:lightcurve} at $\sim$5780~s, where a decline in
brightness is evident. The decline is more gradual than that observed
at blue wavelengths in \citet{bc91}, but is at a similar phase.
\citet{bc91} state that the duration of the bright phase, when spot~1
is in view, is consistent with the latitude and longitude they derive
for the spot for $q=0.2$. This is true only if the optically emitting
region is a small height above the surface of the white dwarf
($0.025R_{\rm WD}$), as expected for the standard accretion scenarios
(see \citealt{cwr+99}). The decline in brightness at $\sim$5780~s in
Figure~\ref{fig:lightcurve} is also consistent with the latitudes and
longitudes we derive in Table~\ref{tab:regions}, for all values of $q$
tabulated, if we assume a height of $0.02-0.025R_{\rm WD}$ (a more
precise height determination is prevented by the gradual nature of the
decline in our data). The height of the region depends in an inverse
manner on the local accretion rate. The more gradual decline in our
data may therefore be a result of a slightly lower height caused by
the higher accretion rate in the intermediate state. The smaller drop
in brightness compared to the spot brightness at eclipse egress can be
qualitatively explained by the beaming of the cyclotron radiation and
by an increase in the contribution of the accretion stream as it is
viewed increasingly side-on.

It may be questioned whether the decline at $\sim$5780~s is in fact the
signature of spot~1 rotating over the limb. In order to remain in
view, the consequence would be that the height of the optically
emitting region must be significantly higher, requiring a change in
spot latitude and a different inclination from those derived in
Table~\ref{tab:regions} and in \citet{bc91}. It is also questionable
whether a small localised region of emission, consistent with both
ingress and egress spot eclipse durations, is physically possible at
heights far above the white dwarf. We therefore prefer the
interpretation that the decline at $\sim$5780~s is due to the rotation
of spot~1 over the limb.

We conclude the following: in the low state there is no evidence for
spot~2 or an accretion stream to spot~1; in the intermediate state
accretion begins at spot~2 and some stream becomes evident (whether to
spot~1 or 2 must be subject to more detailed modelling); in the faint
high and bright high states the accretion increases at spot~1, and the
stream becomes brighter still. This indicates that in the optical the
stream brightness is strongly dependent on the accretion rate -- more
so than the spot brightness.

It is also evident that the emission from spot~2 is greatest in the
intermediate state and that it does not increase commensurately with
the emission from spot~1 at higher accretion rates. In the higher
states at the time of the {\it EUVE\/} observations (\citealt{wsv95})
there is no evidence for spot~2 in the EUV. In the intermediate state
when spot~2 is clearly visible in the optical, there are no detectable
X-rays in the {\it ROSAT\/} soft or hard bands (\citealt{bai95}).

While the reduction in emission from spot~2 in the brighter states can
be explained by changes in the accretion stream trajectory at higher
accretion rates, it is difficult to explain the absence of soft or
hard X-rays from spot~2 (\citealt{bai95}) in the intermediate state.
One possibility could be that the foot of the accretion region is {\it
behind\/} the limb of the primary at the time it is eclipsed, with
only the top (optical cyclotron emitting) of the postshock accretion
flow visible above the limb. In order to be consistent with the
phasing of spot~2 ingress and egress, the co-latitude would be limited
to only a few degrees more than $90^{\circ}-i$, the longitude must be
$180^{\circ}$: no solution is obtainable for the $q=0.3$ case, and it
is marginal for the $q=0.2$ case. For $q=0.15$ a solution with a
co-latitude of $11^{\circ}$ and an emitting height above the white
dwarf surface of $0.025\,R_{\rm WD}$ is consistent. This has the
advantage of naturally explaining the disappearance of the second spot
at higher accretion rates (when the post-shock region cools more
efficiently and is lower). On the other hand we have argued in
Section~\ref{sec:geometry} that mass ratios in the range $0.2-0.3$ are
to be preferred if we accept the constraints of the secondary star
period-mass relationships, and the special conditions of a grazing
occultation of the accretion column may be considered unlikely.

We suggest instead that the accretion rate at spot 2 is low enough
that no accretion shock forms: instead the accreting material
interacts with the white dwarf surface layer directly (the
`bombardment solution' of \citealt{kp82}). This has been investigated
for parameters appropriate to UZ For by \citet{wb92}, with the
prediction that the optical flux produced by cyclotron radiation is
sufficient to produce the observed spot brightness ($\sim 0.05$ mJy).
At the same time the temperature in the optically thick atmospheric
layers emitting X-rays is $\sim10$ eV, too low to be detected in the
{\it ROSAT\/} band, even for the low interstellar absorption to this
system (\citealt{rmc+94}).

\section{Conclusions}

The high time resolution, quantum efficiency, dynamic range and
intrinsic colour resolution of the new STJ detector has made it
possible to explore the accretion processes in even faint magnetic
cataclysmic variables with new levels of precision. The
photon-counting nature of the instrument allows the data to be
combined flexibly, {\it a posteriori}, in spectral and temporal bins
limited only by the photon statistics, while the limited array size is
nevertheless sufficient to provide simultaneous sky subtraction.

We have obtained data for three eclipses of UZ~For. We attribute two
sharp changes in brightness to the eclipse of two small accretion
regions and localise them on the surface of the white dwarf primary.
The first of these is at the lower hemisphere at the location seen by
others in the optical (e.g.\ \citealt{bc91}), and in the EUV and X-rays
(\citealt{wsv95}; \citealt{bai95}). The second is in the upper
hemisphere, near the rotation axis, and there is no evidence for any
emission from this region in X-rays. We have explained this in terms of
the bombardment solution of Kuijpers \& Pringle (1982) which produces
optical emission from a thin interaction layer, while requiring the
X-ray emission to be sufficiently cool to be unobservable by {\it
ROSAT\/}. However, we also have to consider the possibility of a
grazing occultation of this region, although the special conditions
required are somewhat unsatisfactory.

We have updated the orbital ephemeris using those timings we believe
to be of the highest accuracy. This will be important for future
investigations of the spot longitude changes and correlations with
accretion rate. Light time corrections to the Solar System barycentre
are required to obtain timing accuracies below $\sim$3~s, which is
significant in comparison with the diameter of the white dwarf of
$\sim$40~s.

The accretion regions are resolved in our data, and we derive a spot
size as a fraction of the white dwarf surface of slightly less than
$10^{-3}$, somewhat smaller than the size usually derived for
accretion spots in optical data. This may result partly from a lower
time resolution and S/N ratio in other data compared to that available
here.

We have looked for, and placed strong upper limits on, QPOs on
time-scales of order 1~s, as seen in some other AM~Her systems. We
conclude that the magnetic field and thus cyclotron emission is
sufficient to stabilise the post-shock flow (\citealt{sw99}).
 
An analysis of the accretion spots and the accretion stream
brightness, along the lines of that in \citet{hch+99}, is planned for a
subsequent paper. Significantly higher spectral resolution STJ data
expected in the future would be well matched to obtain low-resolution
cyclotron spectra in AM~Her systems. By co-adding several eclipses, it
will be possible to obtain this spectral data even for short phase
intervals, e.g.\ after eclipse ingress of spot~1 and before that of
spot~2. In brighter systems a reconstruction of the accretion spot
will be possible in several spectral bands.

\section*{Acknowledgements}
We acknowledge the contributions of other members of the Astrophysics
Division of the European Space Agency at ESTEC involved in the optical
STJ development effort, in particular J.~Verveer and S.~Andersson (who
also provided technical support at the telescope) and P.~Verhoeve for
the evaluation of device performance. We acknowledge D.~Goldie,
R.~Hart and D.~Glowacka of Oxford Instruments Thin Film Group for the
fabrication of the array. The valuable support of B.~Christensen and
J.~O'Leary on the camera software development is also acknowledged. We
are grateful for the assignment of engineering time at the William
Herschel Telescope of the ING, and we acknowledge the excellent
support given to the instrument's commissioning, in particular by
P.~Moore and C.R.~Benn. We are also grateful to K.~Wu for useful
discussions, to D.~O'Donoghue for the continuing use of his DFT
software, and to L.~Angelini for advice on use of the {\sevensize
XRONOS} package in {\sevensize FTOOLS}. We are grateful to the referee
for identifying shortcomings in our original discussions in
Section~3.7.


\bibliographystyle{mn2e}	


\bsp
\end{document}